\documentclass[10pt, aps, prd, amsmath, floats,floatfix,twocolumn, superscriptaddress]{revtex4-2}

\def\sideremark#1{\ifvmode\leavevmode\fi\vadjust{\vbox to0pt{\vss% the remark
 \hbox to 0pt{\hskip\hsize\hskip1em%                          will appear only

 \vbox{\hsize3cm\tiny\raggedright\pretolerance10000%          on the side
 \noindent #1\hfill}\hss}\vbox to8pt{\vfil}\vss}}}%

\usepackage{mathrsfs}
\usepackage{natbib}
\usepackage{amsmath}
\usepackage{amssymb}
\usepackage{lmodern}

\usepackage[T1]{fontenc}
\usepackage[latin9]{inputenc}
\usepackage{geometry}
\usepackage{subfigure,lmodern, amsmath,amssymb, graphicx, pifont, adjustbox, bm, xcolor}
\usepackage{amsfonts}
\usepackage{geometry}
\usepackage{comment}
\usepackage{mathtools}
\usepackage{slashed}
\usepackage{ragged2e}

\usepackage{bm}

\newcounter{mnotecount}%[section]

\newcommand{\mnotex}[1]%{}
{\protect{\stepcounter{mnotecount}}$^{\mbox{\footnotesize $\bullet$\themnotecount}}$ 
\marginpar{%\color{red}%
\raggedright\tiny\em
$\!\!\!\!\!\!\,\bullet$\themnotecount: #1} }

\begin{document}

\title{A Breathing Universe is Consistent}

\author{Samuel Blitz}
 \email{blitz@math.muni.cz}
  \affiliation{Department of Mathematics and Statistics, Masaryk University, Brno, Czech Republic} 
  
\date{\today}
\begin{abstract}

\noindent

We consider a toy FRW universe with the exotic topology $S^1 \times S^3$. We show that for a specific choice of quantum field content, the Friedmann equations arising from semi-classical general relativity are consistent with temporal periodicity as required by the $S^1$ timelike factor. A straightforward consequence is that entropy reversals occur during each cycle, consistent with Hawking's proposed connection between the thermodynamic and cosmological arrows of time.

\end{abstract}

\maketitle

\noindent

\section{Introduction}
According to the $\Lambda$CDM model, the final fate of the cosmos will likely be one of heat death---an eternal future of exponential expansion governed by a positive cosmological constant in a maximum entropy state~\cite{AdamsLaughlin,KraussStarkman}. But recently, the $\Lambda$CDM model has experienced observational tension in the form of the Hubble tension~\cite{BernalVerdeRiess,Valentino1,Poulin,Valentino2}, the $\sigma_8$ tension~\cite{Planck,DES,Wyman}, and the possible observation of dynamic dark energy~\cite{DESI}. Regardless, however, it is generally accepted that the total entropy of the universe obeys the second law of thermodynamics, barring speculative behavior (like the ekpyrotic universe~\cite{SteinhardtTurok}, conformal cyclic cosmology~\cite{CCC}, or a turnaround cosmology~\cite{Baum}, to name a few).

One notable investigation into the contrary was Hawking's proposed connection between the thermodynamic and cosmological arrows of time~\cite{Hawking1985}---he justifies this claim by examining the CPT invariance of the quantum state of the universe. If we take this proposal seriously, then any expanding universe that recollapses after a finite time will result in a reversed entropic arrow of time. Indeed, if an entropy reversal would indeed occur during a contraction phase, then it may be possible for such a universe to be truly periodic: expanding in one phase and contracting in another, only to repeat itself ad infinitum.

To examine the feasibility of such an idea, it is the goal of this manuscript to establish that there exist \textit{genuinely periodic} solutions to the Einstein field equations in the presence of well-understood quantum fields (even if they are perhaps exotic). Note that quasi-periodic models are known to exist, for example in Brans-Dicke theory~\cite{Barrow} and in the simple harmonic universe~\cite{Graham}. Furthermore, while general relativity is compatible with expansion and recollapse (for example, in a closed FRW spacetime with ordinary fluids), without exotic matter content or modifications to the theory, these collapses inevitably result in a final singularity rather than a bounce off a minimum non-zero size. Thus, true periodicity is challenging to establish without invoking speculative quantum gravitational effects. 

Establishing genuine periodicity implies proving compatibility with an exotic spacetime topology. The simplest such topology that demands genuinely periodic behavior is $S^1 \times M^3$, with $S^1$ a timelike factor. Observe that this compactification is distinct from Kaluza--Klein-like compactifications (see~\cite{Overduin} for a nice review), wherein \textit{spacelike} dimensions are topologically non-trivial. As our universe is expanding, we require one more feature of any such periodic spacetime: that it has an evolving scale factor. It is important to recognize, however, that we do not claim observational viability from this investigation: this work is merely intended as a controlled exercise to explore the possibility of topology-driven periodicity.

To that end, we consider the Lorentzian warped product manifold $S^1 \times S^3$ equipped with the metric
\begin{align} \label{FRW-metric}
ds^2 = -dt^2 + a(t)^2 d \Omega_3^2\,,
\end{align}
where $d \Omega_3^2$ is the metric on the unit 3-sphere, $t \in [0,\tau)$, and $a(t)$ is periodic with period $\tau$; we call such a spacetime a \textit{breathing universe}. In such a universe, every timelike geodesic intersects every separating spatial hypersurface infinitely many times, indicating that closed timelike curves are the norm. These closed timelike curves of course entail the usual pathologies for test particles; however, the matter content in our construction will necessarily obey consistency conditions. That said, so long as $a(t)$ never vanishes, such a universe is geodesically complete as it is a compact manifold with a conformal Killing vector $a \partial_t$, see~\cite{RomeroSanchez}. Reinforcing Hawking's proposal, then, is the recent observation that quantum systems occupying closed timelike curves undergo entropy reversals~\cite{Gavassino}.  We will use a semi-classical gravitating framework to establish the consistency of some such universes.

The layout of this manuscript is as follows. In Section~\ref{sec:energy}, we provide an account of the kinds of field content that may arise in such a universe and describe the energy density of possible states that might arise therein. In Section~\ref{sec:friedman}, we use the semi-classical Einstein field equations to demonstrate that a scale factor in a candidate FRW metric with the aforementioned field content is necessarily periodic, establishing consistency with the topology and behavior of a breathing universe. In Section~\ref{sec:entropy} we briefly speak on the entropic consequences of such a spacetime filled with the specified field content, and in Section~\ref{conclusion} we conclude with some remarks.

\section{Energy Density in a Breathing Universe} \label{sec:energy}
In order to describe the stress-energy tensor for a quantum field theory in a breathing universe, we must first discuss the states that may exist there. In time-dependent spacetimes with Cauchy hypersurfaces, one may build a vacuum state using the adiabatic expansion~\cite{ParkerFulling,LudersRoberts}. However, in a spacetime with a topology as pathological as $S^1 \times S^3$, there are no Cauchy hypersurfaces and so this method fails. Accordingly, we instead formulate the problem as a periodic boundary-value problem on $S^1$ rather than a Cauchy initial value problem. Nonetheless, the existence of regular states in this setting is a nontrivial assumption. Accordingly, we proceed under the semiclassical working hypothesis that there exists a spatially homogeneous and isotropic state on $S^1 \times S^3$ whose stress-energy tensor satisfies $\nabla^a \langle T_{ab} \rangle =0$, the renormalized trace takes the standard anomaly form, and $\langle T_{ab} \rangle$ is of perfect-fluid form by FRW symmetry.

As any quantum state couples to the geometry of the universe (which by construction must evolve periodically), we will model the matter sector as a periodically driven system with period $\tau$. In many open Floquet settings~\cite{Mori,Santoro}, the long-time asymptotic state is a $\tau$-periodic steady state well approximated by the Floquet--Gibbs mixed state
$$\varrho = e^{-\tau \hat{H}_F}/\operatorname{Tr}(e^{-\tau \hat{H}_F})\,,$$
where $\tau$ is the period of $S^1$ and $\hat{H}_F$ is the Floquet Hamiltonian~\cite{FloqExplanation}, defined according to
$$U(\tau,0) = e^{-i \tau \hat{H}_F}\,,$$
with $U(\tau,0)$ the unitary time evolution operator for a full period of the steady state. Recall that so long as $U(\tau,0)$ is unitary, $\hat{H}_F$ is self-adjoint~\cite{Schnell}. We assume we are in such a regime, and hence may use the resulting steady state as our effective matter content and $\hat{H}_F$ as its self-adjoint Floquet Hamiltonian. Under this assumption, the quantum state is a well-defined and thermal KMS state~\cite{Kubo,MartinSchwinger}; here, KMS refers merely to the equilibrium condition with respect to the Floquet evolution generated by $\hat{H}_F$. And as we will be working in a semi-classical setting, we expect that the relevant quantum states would be mixed states---after all, we are essentially tracing out the quantum spacetime degrees of freedom. We thus use this state to describe our matter content.

With these semiclassical assumptions in place, we now turn to the energy density. Now recall that on an FRW spacetime, any homogeneous and isotropic stress-energy tensor has perfect-fluid form at the level of tensor structure~\cite{HawkingEllis}, and hence the stress-energy tensor for our Floquet--Gibbs state is determined by an energy density, a 4-velocity, and an isotropic pressure (which may incorporate viscous effects). As $\varrho$ is defined according to the canonical timelike direction (determined by the $S^1$ geometry), we may choose the 4-velocity to be proportional to $\partial_t$.

Now for a general quantum state, there are generally two (scheme-independent) contributions: one contribution that is state-dependent and one contribution that is state-independent. Computing the former will require a description of the behavior of Floquet--Gibbs thermal states and the latter will arise from the celebrated trace anomaly~\cite{Duff}.

As the quantum state is thermal, its energy density is well-studied~\cite{Altaie}. Now, the model under consideration is a toy model for our long-lived universe, so we only concern ourselves with the energy density in the long period limit (corresponding to low Floquet temperature $1/\tau$):
$$\rho_{thermal} \propto a(t)^{-4}\,.$$
This is the Casimir energy of the system, which is well-defined as spatial slices are compact. Note that for large but finite periods, this energy density will have additional  contributions that depend on $a(t)$ but they are exponentially suppressed in the (large) ratio $\tau/a(t)$.

One may also consider the case where the universe has a short period, corresponding to a high temperature. In that case, the first contributing factor to the thermal energy density is the Stefan--Boltzmann-like term which goes as $\tau^{-4}$ and is independent of the scale factor. This could play the role of a cosmological constant. Higher order corrections pick up scale factor dependence, for example scaling as $\tau^{-2} a^{-2}$. These corrections significantly complicate the dynamics compared to the long-lived breathing universe, and perhaps such a model is worthy of independent study. Nonetheless, we do not consider this here.

So the state-dependent component of the energy density is merely the Casimir energy of the field content on a manifold with topology $S^1 \times S^3$. For massive (with dimensionless mass $\mu$) dimension-1 conformally-coupled scalar fields, the Casimir energy is approximately given by
$$E^{1,\mu \ll 1} _{cas} \approx \frac{1}{240 a} - \frac{\mu^2}{48a}$$
in the small $\mu$ limit and is exponentially suppressed to zero in the large $\mu$ limit~\cite{ElizaldeTort1,ElizaldeTort2}. On the other hand, the Casimir energy for a massless dimension-0 conformally-coupled scalar field is given by~\cite{ArosBugini1,ArosBugini2}
$$E^{0,\mu = 0}_{cas} = -\frac{1}{30a}\,.$$
Similar results can be established for other types of field content, but we will find that these types of field content are sufficient for our purposes, as our goal is merely to show that there exist semi-classical spacetimes with well-understood quantum fields that are consistent with genuine periodicity.

So, when the field content consists of $n_0$ massive (with identical masses) dimension-1 conformally-coupled scalar fields and $n_0'$ massless dimension-0 conformally-coupled scalar fields, the Casimir energy density is given by
$$\rho_{cas} = \frac{n_0 E^{1,\mu}_{cas}}{2 \pi^2 a^3} - \frac{n_0'}{60 \pi^2 a^4}\,.$$
In the large $\mu$ limit (which we adopt going forward), we have that
$$\rho_{cas} \approx - \frac{n_0'}{60 \pi^2 a^4} < 0\,.$$

Finally, in a curved spacetime, quantum fields produce state-independent contributions to the energy density that are both independent of the renormalization scheme chosen and are 1-loop exact: these contributions are captured by the trace anomaly~\cite{Duff,BoyleTurok}:
$$\langle T_a^a \rangle =  - \alpha E_4 + \gamma C_{abcd} C^{abcd}\,,$$
where $C_{abcd}$ is the Weyl tensor,
$$E_4 := R^{abcd} R_{abcd} - 4 R^{ab} R_{ab} + R^2$$
is the Euler density, and $\alpha$ and $\gamma$ are constant coefficients that depend on the field content. These two terms, $E_4$ and $|C|^2$, are often called type-A and type-C anomalies, respectively. Note that there is also a contribution to the trace of the stress-energy tensor present for quantum fields that does depend on the renormalization scheme---a total derivative term of the form $\square R$. In particular, one may choose the dimensional continuation and (modified) minimal subtraction so that the coefficient of $\square R$ is set to zero in the renormalized trace while keeping the left-hand side in Einstein--Hilbert form (i.e. no finite curvature-squared terms)~\cite{Duff,Burgess,AGS}. This is a standard technique~\cite{mehdizadeh}, so we adopt this scheme for simplicity.

As an FRW spacetime is conformally-flat, the type-C anomaly vanishes, and we are only left with the type-A anomaly. In particular, for a field theory with $n_0$ dimension-one scalars and $n'_0$ dimension-zero scalars, the coefficient $\alpha$ is given by~\cite{BoyleTurok}
\begin{align*}
\alpha &= \tfrac{1}{360 (4 \pi)^2} \left(n_0 - 28 n'_0 \right)\,.
\end{align*}
By judiciously choosing the particle content of the field theory, one may select $\alpha$ to have either sign and the magnitude is unbounded.

Now, in a breathing universe with a metric given by Equation~(\ref{FRW-metric}), we can explicitly compute the trace anomaly:
$$\langle T_a^a \rangle = -24 \alpha (H^2 + a^{-2})(\dot{H} + H^2)\,.$$
Recalling that $\dot{H} + H^2 = \ddot{a}/a$, it follows that
\begin{align} \label{trace-anomaly}
\langle T_a^a \rangle = -\frac{24 \alpha\, \ddot{a}(\dot{a}^2 + 1)}{a^3}\,.
\end{align}

\medskip

To determine the semi-classical stress-energy tensor using the trace anomaly and the energy densities arising from the Casimir energy, we need to invoke the continuity equation and the fact that, on an FRW metric, the stress-energy tensor of a homogeneous source is that of a perfect fluid at the level of tensor structure. More specifically, it must be the case that 
$$\langle T_a^a \rangle = -\rho + 3 p\,.$$
Note that we do not assume an equation of state---on FRW spacetimes, symmetry implies that the matter content is an effective isotropic fluid, and the trace anomaly supplies $\langle T_a^a \rangle$, and hence $p$ is determined. Now solving the above for $p$ and substituting into the continuity equation
$$\dot{\rho} + 3 H (\rho + p) = 0\,,$$
we have on the FRW background that
$$\dot{\rho} + 4H \rho + H \langle T_a^a \rangle = 0\,.$$
We can solve the differential equation, yielding
\begin{align} \label{energy-density}
\rho = 6\alpha (H^2 + a^{-2})^2 + \frac{C}{a^4}\,,
\end{align}
where $C$ is an integration constant.  As energy densities of the form $C/a^4$ for an effective isotropic source do not contribute to the trace of the stress-energy tensor, we identify $C := - \frac{n_0'}{60 \pi^2}$ with the state-dependent part of the energy density associated with the specific state of the quantum system: in this case, the low-temperature thermal state that has a Casimir energy. Thus, the energy density may then be viewed as arising from two distinct components: the first term on the right side of Equation~(\ref{energy-density}) is the state-independent contribution from the trace anomaly, while the second term is the state-dependent Casimir/thermal contribution of the particle fields. In particular, in the large-$\mu$ limit adopted here, the Casimir/thermal term is negative, whereas for sufficiently large $n_0$ and $n_0'$, the coefficient $\alpha$ may be positive, so the sign of the total energy density need not agree with the sign of the Casimir contribution alone. Thus the total semiclassical energy density may remain positive even though the Casimir part is negative. Note that the atypical scaling for a Casimir energy is expected: Casimir energies in $S^3$ scale as $a^{-4}$~\cite{Altaie}. This will be critical in the next section.

\section{A Periodic Scale Factor is Possible} \label{sec:friedman}
We now consider the semi-classical Einstein field equations sourced by $n_0$ (very) massive dimension-1 conformally-coupled scalar fields and $n_0'$ massless dimension-0 conformally-coupled scalar fields in the thermal state described in the previous section. The semi-classical Einstein field equations are
$$G_{ab} + \Lambda g_{ab} = 8 \pi G \langle T_{ab} \rangle\,,$$
where $\langle T_{ab} \rangle$ is the effective isotropic fluid stress-energy tensor consistent with the energy density in Equation~(\ref{energy-density}) and $\Lambda$ is the (possibly non-zero) cosmological constant. Our goal in this section is to show that there exists a solution to this equation on $S^1 \times S^3$ for some choice of the cosmological constant $\Lambda$ and field content $(n_0,n_0')$.

Now supposing that the metric takes the form in Equation~(\ref{FRW-metric}) with spatial slices having topology $S^3$, the first Friedmann equation follows:
\begin{align} \label{friedman-dynamics}
H^2 + a^{-2} = \frac{8 \pi G \rho}{3} + \frac{\Lambda}{3}\,,
\end{align}
where $\rho$ is the energy density in Equation~(\ref{energy-density}). Our goal is now to study this differential equation.

If this Friedmann equation has periodic solutions, then turning points occur when $H = 0$. As the scale factor must be real and non-negative, the only candidate turnaround points occur at
$$a_{\pm} = \sqrt{\frac{3 \pm \sqrt{9 - 32 \pi G \Lambda (6 \alpha + C)}}{2 \Lambda}}\,,$$
with $a_- < a_+$.

%Defining $u = H^2 + a^{-2}$, this system becomes
%$$u = 16 \pi G \alpha u^2 + \left(\frac{\Lambda}{3} + \frac{8 \pi G C}{3a^4} \right)\,.$$
%This quadratic equation can be solved, yielding
%$$u_{\pm} = \frac{1 \pm \sqrt{1 - 64 \pi G \alpha \left(\tfrac{\Lambda}{3} + \frac{8 \pi G C}{3a^4}\right)}}{32 \pi G \alpha}\,.$$
%So, the Friedman equation reduces to
%\begin{align} \label{friedman-dynamics}
%H^2 = \frac{1 \pm \sqrt{1 - 64 \pi G \alpha \left(\tfrac{\Lambda}{3} + \frac{8 \pi G C}{3a^4}\right)}}{32 \pi G \alpha} - \frac{1}{a^2}\,.
%\end{align}
%
%If this Friedman equation has periodic solutions, then turning points occur when $H = 0$. (We must also have that $H^2 \geq 0$ everywhere---this will be verified after we have established what the required range of parameters are for a cyclic solution.) On the positive branch of Equation~(\ref{friedman-dynamics}), we find a pair of extrema $a_- < a_+$ given by
%$$a_{\pm} = \sqrt{\frac{3 \pm \sqrt{9 - 32 \pi G \Lambda (6 \alpha + C)}}{2 \Lambda}}\,.$$
%It turns out that the negative branch has the same solutions, so we only consider the positive branch going forward. 
Now, in order for these to be real-valued extrema, respectively, we require that $\Lambda > 0$, and for
$$g := 9 -32 \pi G \Lambda (6 \alpha + C)\,,$$
we must have that $0 < g < 9$.

To establish that these extrema are turnaround points, we must check that $\ddot{a}_{-} > 0$ and that $\ddot{a}_{+} < 0$. From the second Friedmann equation, we have that
$$\frac{\ddot{a}}{a} = \frac{\Lambda}{3} - \frac{4 \pi G}{3} (\rho + 3 p)\,.$$
Note that at both extremes, we have that $\rho = (C + 6 \alpha) a^{-4}$ as $\dot{a} = 0$ there. Furthermore, at the same scale factors, we have that
$$\langle T_a^a \rangle|_{\dot{a} = 0} = -\frac{24 \alpha \ddot{a}}{a^3}\,,$$
implying (from the effective isotropic fluid description of the stress-energy) that
$$\frac{\ddot{a}}{a} = \frac{\Lambda}{3} - \frac{4 \pi  G}{3} \left[ 2 (C + 6 \alpha) a^{-4} -24 \alpha \ddot{a} a^{-3} \right]\,.$$
Rearranging this equation, we find that
$$12(a^2 - 32 \pi G \alpha )\Lambda  a \ddot{a} = 4a^4 \Lambda^2 - 32 \pi G \Lambda (C + 6 \alpha) \,.$$
Evaluating this equation on the extrema $a_{\pm}$ and rearranging, it follows that
$$a_{\pm} \ddot{a}_{\pm} = \frac{g \pm 3 \sqrt{g}}{32 G \Lambda \pi C + g \pm 3 \sqrt{g} }\,.$$
It is now straightforward to check that when
$$C < - \frac{g + 3 \sqrt{g}}{32 \pi G \Lambda}\,,$$
the turnaround conditions given by $\ddot{a}_- > 0$ and $\ddot{a}_+ < 0$ are both satisfied.

The parameter constraints can only be satisfied for certain values of $(\alpha,C)$.
Importantly, it is not the case that for \textit{any} choice of $\alpha$, there exists a $C$ satisfying these inequalities. Indeed, examining the constraints more carefully,  we find that they are equivalent to the system
\begin{align*}
\frac{3}{64 \pi G \Lambda} &< \alpha \\
\operatorname{max}(-128 \pi G \Lambda \alpha^2 + 6 \alpha, -6 \alpha) &< C < -\frac{3(64 \pi G \Lambda \alpha - 3)}{32 \pi G \Lambda}\,.
\end{align*}
Thus, for the range of values $(\alpha, C)$ allowed by the above inequalities, genuine periodic behavior is possible for the scale factor of an FRW metric with spatial slices given by $S^3$. To establish that periodicity indeed happens, we note that by defining $u := H^2 + a^{-2}$, the Friedmann equation may be written as
$$16 \pi G \alpha u^2 - u + \frac{\Lambda}{3} + \frac{8 \pi G C}{3a^4} = 0\,.$$
Solving for $u$ and keeping the branch whose equilibrium solutions are $a_{\pm}$, we obtain
$$u = \frac{1 + \sqrt{1 - \tfrac{64 \pi G \alpha}{3} \left(\Lambda + \tfrac{8 \pi G C}{a^4} \right)}}{32 \pi G \alpha}\,.$$
Solving for $\dot{a}^2$ using the definition of $u$, we have that 
$$\dot{a}^2 = \frac{3a^2 - 96 \pi G \alpha + a^2 \sqrt{9 - 192 \pi G \alpha \left(\Lambda + \tfrac{8 \pi G C}{a^4} \right)}}{96 G \pi \alpha}\,.$$

To establish periodic behavior, it suffices to show that the right hand side of the above equation is positive in the interval $(a_-,a_+)$ for the allowed parameters. As $\Lambda > 0$ and thus, from the constraints, we require that $\alpha > 0$, we must only show that
\begin{align} \label{positivity}
F(a):=3a^2 - 96 \pi G \alpha + a^2 \sqrt{9 - 192 \pi G \alpha \left(\Lambda + \tfrac{8 \pi G C}{a^4} \right)} > 0\,.
\end{align}
Given the allowed parameters, it is easy to show that for all $a_- < a < a_+$, $F(a)$ is real. Furthermore, differentiating $F(a)$, we see that there exists only a single extremum that is positive, occuring at
$$a_{mid} := \sqrt[4]{-\frac{24 \pi G C}{\Lambda (64 \pi G \Lambda \alpha - 3)}}\,.$$
Differentiating again and evaluating on $a_{mid}$, we find that $a_{mid}$ is a maximum. It follows that $F(a)$ is positive for all $a$ between $a_-$ and $a_+$, which establishes periodic behavior between these two bounds.

\begin{figure}[h]
\includegraphics[width=5cm, height=4cm]{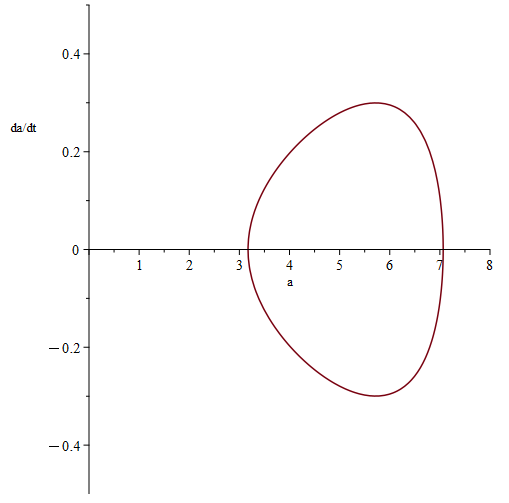}
\caption{An example phase portrait for the Friedmann equation~(\ref{friedman-dynamics}) with parameters $G = 1$, $\Lambda = 0.05$, $\alpha = 1$ and $C = -5$. These parameters satisfy the constraint equations. The presence of a closed orbit indicates periodicity in $a(t)$. Numerically solving the second Friedmann equation with these parameters yields a period of $\tau \approx 43$ and a maximum scale factor of $a_+ \approx 7.1$, meaning the dimensionless ratio characterizing the energy density regime (long vs. short period) for $\rho$ is approximately $6$, which places it well within the long period regime as then $e^{-\mathcal{O}(\tau/a)}$ is small.}
\label{fig:phase-portrait}
\end{figure}

By examining the phase portrait for the dynamic system Equation~(\ref{friedman-dynamics}), we can also easily see periodicity for particular values of $(\alpha, C)$; see Figure~\ref{fig:phase-portrait}. Following the above argument, we may thus identify the time coordinates $t = 0$ and $t = \tau$, where $\tau$ is the period of one cycle. That is, the solution to the semi-classical Einstein field equations is consistent with a breathing universe model. Furthermore, the periodic scale factor $a(t)$ does not necessarily approach the quantum gravity regime (or a singularity) where the semi-classical Einstein field equations would fail---rather, the minimum scale factor $a_-$ is controlled by $\Lambda$ and $g$: smaller $\Lambda$ implies larger $a_-$, while if $g$ approaches $9$, then $a_-$ approaches zero.

So it suffices to check that there exists a cosmological constant $\Lambda$ and field content specified by the positive integers $(n_0, n_0')$ that determines $\alpha$ and $C$ consistent with these inequalities. Observe that for any $n_0'$, we may choose $n_0$ sufficiently large so that the $\alpha$ constraint is satisfied, and indeed so that $\alpha$ is arbitrarily large. For sufficiently large $\alpha$, we find that the $C$ constraint becomes
$$\frac{n_0}{44} - \frac{135 \pi}{22 G \Lambda}< n_0' < \frac{n_0}{44}\,.$$
When $n_0$ is sufficiently large, the left and right extremes above are positive, and so long as $\frac{135 \pi}{22 G \Lambda} > 1$, there exists an integer $n_0'$ between the two extreme constraints. Hence, for any $0< \Lambda < \frac{135 \pi}{22 G}$, there exists some pair $(n_0,n_0')$ satisfying the constraints as required for $a(t)$ to be periodic.

As the exponentially suppressed terms in the energy density arising from corrections to the Casimir energy density can be treated as small perturbations, they do not spoil the periodicity. Their dominant effect is to adjust the minimum scale factor $a_-$, decreasing $a_-$ for negative energy contributions and increasing it for positive energy contributions. Future investigations into the details of the perturbative behavior of such terms may be worthwhile.

It is also interesting to point out that, for the constraints on the parameter space given, the energy density $\rho(t)$ is strictly positive over the entire cycle. This follows because the energy density is minimized at the extrema, so positivity of $\rho(t)$ requires that $C > -6 \alpha$. But this follows trivially from the constraint equation on $C$. However, the weak energy condition is violated: we may explicitly verify that, when $a = a_-$, $\rho + p < 0$. But this is not unexpected, and indeed it is known that quantum systems violate the weak energy condition.

\section{Entropic Consequences of a Periodic FRW Universe} \label{sec:entropy}
We have now established that it is at least plausible that there are self-consistent semi-classical breathing universes. Assuming the matter sector occupies the $\tau$-periodic steady state as described above, \textit{all} expectation values of observables are $\tau$-periodic. The behavior of entanglement entropies is subtler; nevertheless, one can give a simple heuristic for an entropy reversal at the background level, occuring exactly when the contraction phase begins.

Generally speaking, computing the entanglement entropy for any subregions of this system is non-trivial. However, because even the exotic dimension-0 scalar has a local action, familiar arguments~\cite{CasiniHuerta} ensure that, for a given UV cutoff, the entanglement entropy between a subregion $\mathcal{S}$ of the $S^3$ and its complement has leading term proportional to the area of the bounding set $\partial \mathcal{S}$. However, observe that for any choice of $\mathcal{S}$, the area of $\partial \mathcal{S}$ scales as $a(t)^2$. The implication is that, starting when $a(t) = a_+$, the leading area-law contribution to the entanglement entropy of \textit{every} subregion reverses. Of course, this concerns the leading UV-sensitive area term at fixed physical cutoff; subleading state-dependent contributions may modify the detailed entropy evolution. Nonetheless, exact return to the $t = 0$ state is required at $t = \tau$.

At the isotropic background level, we have thus established that Hawking's proposal survives in a breathing universe: the cosmological arrow of time matches the thermodynamic arrow of time. However, Page pointed out~\cite{Page} that Hawking's treatment of CPT invariant universal wavefunctions do not necessarily imply entropy reversal upon collapse: rather, the evolution of entropy also depends on the behavior of dynamics governed by the growth of density perturbations, which do not necessarily align with the background scale factor evolution. In a breathing universe model, these perturbations must in fact be periodic over each cycle. We expect this periodicity to impose additional constraints on the allowed parameter space, but a full perturbation analysis is beyond the scope of this work and is left for future investigation.

\section{Conclusion} \label{conclusion}
In this manuscript, we have established that there exists a family of solutions to the semi-classical Einstein field equations that results in a breathing universe. A bizarre feature of this family of universes is that, in some sense, such a universe has a finite age despite not having a ``beginning'' and indeed being geodesically complete. More specifically, there is an upper bound on the time recordable by a comoving stopwatch regardless of the time coordinates at which it starts and stops. This is in contrast with the $\Lambda$CDM model, where a future heat death ensures that a clock can record any finite value of time.

To be sure, our universe is not such a simple universe: our universe is much more complicated than the breathing universe discussed here and the scale factor behavior does not match observational data. Nonetheless, such solutions to the semi-classical Einstein field equations established here are novel and demonstrate that, even without adding speculative quantum gravitational effects or modified gravitational theories, it is at least \textit{possible} that our universe may have a similarly exotic cyclic topology.

\begin{acknowledgements}
The author would like to acknowledge Jaroslaw Kopinski, Gabriel Herczeg, and Robert Scherrer for their insightful comments during preparation of this manuscript.
\end{acknowledgements}


\begin{thebibliography}{10}
\bibitem{FloqExplanation} The Floquet Hamiltonian is the effective time-independent Hamiltonian that reproduces the stroboscopic evolution of a periodically driven system, i.e. the evolution from one full cycle to the next.
\end{thebibliography}
\end{document}